\documentclass{aa}
\usepackage{txfonts}
\usepackage{graphicx}
\usepackage{subcaption}
\usepackage[colorlinks=true,allcolors=blue]{hyperref}

\begin{document} 

\titlerunning{}
\authorrunning{Mountrichas et al.}
\titlerunning{}

\title{The link between star-formation and supermassive black hole properties}

\author{George Mountrichas\inst{1} \& V$\rm \acute{e}$ronique Buat\inst{2,3}}
          
     \institute {Instituto de Fisica de Cantabria (CSIC-Universidad de Cantabria), Avenida de los Castros, 39005 Santander, Spain
              \email{gmountrichas@gmail.com}
              \and
             Aix Marseille Univ, CNRS, CNES, LAM Marseille, France. 
              \and
                 Institut Universitaire de France (IUF) }

\abstract{It is well known that supermassive black holes (SMBHs) and their host galaxies co-evolve. AGN feedback plays an important role on this symbiosis. To study the effect of the AGN feedback on the host galaxy, a popular method is to study the star-formation rate (SFR) as a function of the X-ray luminosity (L$_X$). However, hydrodynamical simulations suggest that the cumulative impact of AGN feedback on a galaxy is encapsulated in the mass of the SMBH, M$_{BH}$, rather than the L$_X$. In this study, we compare the SFR of AGN and non-AGN galaxies as a function of L$_X$, M$_{BH}$, Eddington ratio (n$_{Edd}$) and specific black hole accretion rate ($\lambda _{sBHAR}$). For that purpose, we use 122 X-ray AGN in the XMM-XXL field and 3371 galaxies from the VIPERS survey  and calculate the SFR$_{norm}$ parameter, defined as the ratio of the SFR of AGN to the SFR of non-AGN galaxies with similar stellar mass, M$_*$, and redshift. Our datasets span a redshift range of $\rm 0.5\leq z\leq 1.2$. The results show that the correlation between SFR$_{norm}$ and M$_{BH}$ is stronger compared to that between SFR$_{norm}$ and L$_X$. A weaker correlation is found between SFR$_{norm}$ and $\lambda _{sBHAR}$. No correlation is detected between SFR$_{norm}$ and n$_{Edd}$. These results corroborate the idea that the M$_{BH}$ is a more robust tracer of the cumulative impact of the AGN feedback compared to the instantaneous accretion rate (L$_X$) and, thus, a better predictive parameter of the changes of the SFR of the host galaxy.}

   
\maketitle

\section{Introduction}

The supermassive black holes (SMBHs) that live in the centre of galaxies become active  when material that is in the vicinity of the SMBH is accreted onto them. Many evidence have been presented the last two decades that show that there is a co-evolution between the SMBH and its host galaxy. For instance, both the activity of the black hole and the star-formation (SF) of galaxies are fed by the same material (i.e., cold gas) and both phenomena peak at about the same cosmic time \citep[$\rm z\sim 2$; e.g.,][]{Boyle2000, Sobral2013}. Moreover, tight correlations have been found in the local universe, between the mass of the SMBH, M$_{BH}$, and various properties of the host galaxy, such as the stellar velocity dispersion the bulge luminosity and the bulge mass \citep[e.g.,][]{Magorrian1998,Ferrarese2000, Tremaine2002, Haring2004}. These correlations also seem to exist at higher redshifts \citep[$\rm z\sim 2$; e.g.][]{Jahnke2009, Merloni2010, Sun2015, Suh2020, Setoguchi2021, Mountrichas2023b}. 

Various mechanisms have been suggested that drive the gas from kiloparsec to sub-parsec scales \citep[for a review see][]{Alexander2012}.  AGN feedback in the form of jets, radiation, or winds is also included in most simulations to explain many galaxy properties, such as to maintain the hot intracluster medium \citep[e.g.,][]{Dunn2006}, to explain th shape of the galaxy stellar mass function \citep[e.g.,][]{Bower2012} and the galaxy morphology \citep[e.g.,][]{Dubois2016}.

A popular method to study the symbiosis between the AGN and its host galaxy is to examine the correlation between the star-formation rate (SFR) and the power of AGN, using as a proxy for the latter the X-ray luminosity (L$_X$). Most previous studies have found a positive correlation between the SFR and L$_X$ \citep[e.g.,][]{Lanzuisi2017, Masoura2018, Brown2019}, although, no correlation has also been reported \cite{Stanley2015}. However, more information can be gained when we compare the SFR of AGN with the SFR of non-AGN galaxies with similar redshifts and stellar masses, M$_*$, as a function of L$_X$ \citep[e.g.][]{Santini2012, Shimizu2015, Shimizu2017, Florez2020}. In this case, most studies measure what is often call normalized SFR, SFR$_{norm}$, which is the ratio of AGN to the ratio of SF main-sequence (MS) galaxies with similar redshift and M$_*$ \citep{Rosario2013, Mullaney2015, Bernhard2019}. A strong positive correlation has been found between SFR$_{norm}$ and L$_X$ at redshifts up to $\rm z\sim 5$ \citep{Masoura2021, Koutoulidis2022, Pouliasis2022}. However, after minimizing systematics effects that may be introduced in the comparison of the SFR of AGN and non-AGN systems \citep[e.g., due to the different methods that the SFR of the two populations have been calculated, the different photometric selection criteria  that have been applied; for more details see][]{Mountrichas2021c}, a weaker correlation or even absence of correlation is detected between SFR$_{norm}$ and L$_X$, depending on the M$_*$ range \citep[see Fig. 5 in][]{Mountrichas2022b}.

The different trends observed in the SFR$_{norm}-$L$_X$ relation in different M$_*$ regimes, also highlight the importance of M$_*$ in this kind of investigations. There are observational works that have found that the black hole accretion rate ($\rm BHAR \propto L_X$) is mainly linked to M$_*$ rather than SFR \citep{Yang2017}. Moreover, SFR$_{norm}$ appears to be stronger correlated with M$_*$ than with L$_X$ \citep{Mountrichas2022b}. Theoretical studies that used hydrodynamical simulations have also found that that the cumulative impact of AGN feedback on the host galaxy is encapsulated in the mass of the supermassive black hole, M$_{BH}$, and not in L$_X$, both in the local universe \citep{Piotrowska2022} and at high redshifts \citep{Bluck2023}. The fact that the SFR shows a strong link both with M$_*$ and M$_{BH}$ could be due to the underlying M$_*$-M$_{BH}$ relation that has been found to hold up to at least redshift of 2 \citep[e.g.,][]{Merloni2010, Sun2015, Setoguchi2021, Mountrichas2023b}.

In this work, we compare the SFR of X-ray detected AGN with that of non-AGN galaxies as a function of different black hole properties. For that purpose, we use X-ray AGN detected in the XMM-XXL field, for which there are available M$_{BH}$ measurements, and (non-AGN) galaxies from the VIPERS survey that (partially) overlaps with XMM-XXL. We use these two samples to calculate the SFR$_{norm}$ parameter and examine the correlation of SFR$_{norm}$ with the L$_X$, M$_{BH}$, Eddington ratio (n$_{Edd}$) and specific black hole accretion rate ($\lambda _{sBHAR}$). Finally, we discuss our results and describe our main conclusions. Throughout this work, we assume a flat $\Lambda$CDM cosmology with $H_ 0=70.4$\,Km\,s$^{-1}$\,Mpc$^{-1}$ and $\Omega _ M=0.272$ \citep{Komatsu2011}.

\section{Data}
\label{sec_data}

The main goal of this study is to examine how the SFR of X-ray AGN compares with the SFR of non-AGN systems as a function of various black hole properties. For that purpose, we compile an X-ray dataset that comprises of AGN detected in the XMM-XXL field and a control sample of (non-AGN) galaxies which consists of sources observed by the VIPERS survey. The sky area that the two surveys cover (partially) overlaps. Below, we provide a brief description of these two surveys. The (final) AGN and non-AGN samples used in our analysis are described in Sect. \ref{sec_final_samples}. 

\subsection{The XMM-XXL dataset}
\label{sec_data_xxl}

The X-ray dataset used in this work, consists of X-ray AGN observed in the northern field of the {\it{XMM-Newton}}-XXL survey \citep[XMM-XXL;][]{Pierre2016}. XMM-XXL is a medium-depth X-ray survey that covers a total area of 50\,deg$^2$ split into two fields nearly equal in size, the XMM-XXL North (XXL-N) and the XXM-XXL South (XXL-S). The XXL-N dataset consists of 8445 X-ray sources. Of these X-ray sources, 5294 have SDSS counterparts and 2512 have reliable spectroscopy \citep{Menzel2016, Liu2016}. Mid-IR and near-IR was obtained following the likelihood ratio method \citep{Sutherland_and_Saunders1992} as implemented in \citep{Georgakakis2011}. For more details on the reduction of the XMM observations and the IR identifications of the X-ray sources, readers can refer to \cite{Georgakakis2017}.

\subsection{The VIPERS catalogue}
\label{sec_data_vipers}

The galaxy control sample used in our analysis comes from the public data release 2 \citep[PDR-2;][]{Scodeggio2016} of the VIPERS survey \citep{Guzzo2014, Garilli2014}, that partially overlaps with the XMM-XXL field. The observations have been carried out using the VIMOS \citep[VIsible MultiObject Spectrograph,][]{LeFevre2003} on the ESO Very Large Telescope (VLT). The survey covers an area of $\approx$ 23.5\,deg$^2$, split over two regions within the CFHTLS-Wide (Canada-France- Hawaii Telescope Legacy Survey) W1 and W4 fields. Follow-up spectroscopic targets were selected to the magnitude limit i$^\prime=22.5$  from the T0006 data release of the CFHTLS catalogues. An optical colour-colour pre-selection, i.e., {\it{[(r-i)$ > $0.5(u-g) or (r-i)$ > $0.7]}}, excludes galaxies at $z<0.5$, yielding a $>98\%$ completeness for $z>0.5$ and up to $\rm z\sim 1.2$ \citep[for more details see][]{Guzzo2014}. PDR-2 consists of 86,775 galaxies with available spectra. Each spectrum is assigned a quality flag that quantifies the redshift reliability. In all VIPERS papers, redshifts with flags in the range between 2 and 9 are considered as reliable and are those used in the science analysis \citep{Garilli2014, Scodeggio2016}. The above criteria yield 45,180 galaxies within the redshift range spanned by the VIPERS survey (0.5$<$z$<$1.2). This is the same galaxy sample used in \cite{Mountrichas2019} (see their Sect. 2.1).

To add near-IR and mid-IR photometry, we cross-match the VIPERS catalogue with sources in the VISTA Hemisphere Survey \citep[VHS,][]{McMahon2013} and the AllWISE catalogue from the WISE survey \citep{Wright2010}. The process is described in detail in Sect. 2.5 in \cite{Pouliasis2020}. Specifically, the xmatch tool from the astromatch\footnote{https://github.com/ruizca/astromatch} package  was used. xmatch utilizes different statistical methods for cross$-$matching of astronomical catalogues. This tool matches a set of catalogues and gives the Bayesian probabilities of the associations or non-association \citep{Pineau2017}. We only kept sources with a high probability of association ($>68\%$). When one source was associated with several counterparts, we selected the association with the highest probability. 14,128 galaxies from the VIPERS catalogue have counterparts in the near- and mid-IR.

\section{Galaxy and supermassive black hole properties}
\label{sec_analysis}

In the following part of this work, we describe how we obtain measurements for the properties of the sources used in our analysis. Specifically, we present how we measure the SFR and M$_*$ of AGN and non-AGN galaxies, how we calculate the bolometric luminosity (L$_{bol}$), n$_{Edd}$ and $\lambda_{sBHAR}$) of AGN and how the available M$_{BH}$ were estimated. 

\subsection{Calculation of SFR and M$_*$}

For the calculation of the SFR and M$_*$ of AGN host galaxies and non-AGN systems, we apply spectral energy distribution (SED) fitting, using the CIGALE algorithm \citep{Boquien2019, Yang2020, Yang2022}. CIGALE allows the inclusion of the X-ray flux in the fitting process and has the ability to account for the extinction of the UV and optical emission in the poles of AGN \citep{Yang2020, Mountrichas2021a, Mountrichas2021b, Buat2021}.

For consistency with our previous studies \citep{Mountrichas2021c, Mountrichas2022a, Mountrichas2022b, Mountrichas2023}, we use the same templates and parametric grid in the SED fitting process as those used in these previous works. In brief, the galaxy component is modelled using a delayed SFH model with a function form $\rm SFR\propto t \times exp(-t/\tau)$. A star formation burst is included \citep{Malek2018, Buat2019} as a constant ongoing period of star formation of 50\,Myr. Stellar emission is modelled using the single stellar population templates of \cite{Bruzual_Charlot2003} and is attenuated following the \cite{Charlot_Fall_2000} attenuation law. To model the nebular emission, CIGALE adopts the nebular templates based on \cite{VillaVelez2021}. The emission of the dust heated by stars is modelled based on \cite{Dale2014}, without any AGN contribution. The AGN emission is included using the SKIRTOR models of \cite{Stalevski2012, Stalevski2016}. The parameter space used in the SED fitting process is shown in Tables 1 in \cite{Mountrichas2021a, Mountrichas2022a, Mountrichas2022b}. 

CIGALE has the ability to model the X-ray emission of galaxies. In the SED fitting process, the intrinsic L$_X$ in the $2-10$\,keV band are used. The calculation of the intrinsic L$_X$ is described in detail in Sect. 3.1 in \cite{Mountrichas2021a}. In brief, we use the number of photons in the soft ($0.5-2$\,keV) and the hard ($2-8$\,keV) bands that are provided in the \cite{Liu2016} catalogue. Then, a Bayesian approach \citep[BEHR;][]{Park2006} is applied to calculate the hardness ratio, $HR= \frac{H-S}{H+S}$, of each source, where H and S are the counts in the soft and hard bands, respectively. These hardness ratio measurements are then inserted in the Portable, Interactive, Multi-Mission Simulator tool \citep[PIMMS;][]{Mukai1993}  to estimate the hydrogen column density, N$_H$, of each source. A power law with slope $\Gamma=1.8$ for the X-ray spectra is assumed. The value of the galactic N$_H$ is N$_H = 10^{20.25}$\,cm$^{-1}$.

The reliability of the SFR measurements, both in the case of AGN and non-AGN systems, has been examined in detail in our previous works and, in particular, in Sect. 3.2.2 in \cite{Mountrichas2022a}. Finally, we note that the AGN module is used when we fit the SEDs of non-AGN systems. This allows us to uncover AGN that remain undetected by X-rays \cite[e.g.,][]{Pouliasis2020} and exclude them from our galaxy control sample (see Sect. 
 \ref{sec_final_samples}).

\subsection{Calculation of SFR$_{norm}$}

The goal of this study is to compare the SFR of AGN host galaxies with the SFR of non-AGN systems, as a function of various black hole properties. For the comparison of the SFR of AGN and non-AGN galaxies, we use the SFR$_{norm}$ parameter. SFR$_{norm}$ is measured following the process of our previous studies \citep[e.g.,][]{Mountrichas2021c, Mountrichas2022a, Mountrichas2022b}. Specifically, the SFR of each X-ray AGN is divided by the SFR of galaxies in the control sample that are within $\pm 0.2$\,dex in M$_*$ and $\rm \pm 0.075\times (1+z)$ in redshift. Furthermore, each source is weighted based on the uncertainty of the SFR and M$_*$ measurements made by CIGALE. Then, the median values of these ratios are used as the SFR$_{norm}$ of each X-ray AGN. We note that our measurements are not sensitive to the choice of the box size around the AGN. Selecting smaller boxes, though, has an effect on the errors of the calculations \citep{Mountrichas2021c}. The calculation of SFR$_{norm}$ requires both datasets to be mass complete in the redshift range of interest. This requirement is met in the stellar mass range we perform our analysis (see Sect. \ref{sec_final_samples}).

\subsection{Black hole mass measurements}
\label{sec_mbh}

Out of the 2512 AGN in the XXL-N catalogue that have reliable spectroscopy from SDSS-III/BOSS (Sect \ref{sec_data}). 1786 have been classified as broad line AGN (BLAGN1), by \cite{Menzel2016}. A source was classified as BLAGN1 using the full width at half-maximum (FWHM) threshold of 1000\,Km\,s$^{-1}$. \cite{Liu2016} performed spectral fits to the BOSS spectroscopy of these 1786 BLAGN1 to estimate single-epoch virial M$_{BH}$ from continuum luminosities and broad line widths \citep[e.g.,][]{Shen2013}. The details of the spectral fitting procedure are given in Sect. 3.3 of \cite{Liu2016} and in \cite{Shen2013}. In brief, they first measured the continuum luminosities and broad line FWHMs. Then, they used several single-epoch virial mass estimators to calculate M$_{BH}$. Specifically, they applied the following fiducial mass recipes, depending on the redshift of the source: H\,$\beta$ at $\rm z<0.9$, Mg~{\sc ii} at $\rm 0.9<z<2.2$ and   C~{\sc iv} at $\rm z>2.2$. 

Previous studies have shown that single-epoch M$_{BH}$ estimates that use different emission lines, when adopting the fiducial single-epoch mass formula, are generally consistent with each other with negligible systematic offsets and scatter \citep[e.g.,][]{Shen2008, Shen2011, Shen2012, Shen2013}. \cite{Liu2016} confirmed these previous findings. Finally, their M$_{BH}$ measurements have, on average, errors of $\sim 0.5$\,dex, whereas sources with higher SNR have uncertainties of the measured M$_{BH}$ that are less than 0.15\,dex.

\subsection{Bolometric luminosity of the AGN, Eddington ratio and specific black hole accretion rate calculations}
\label{sec_lbol_nedd_lambda}

There are two measurements available for the L$_{bol}$ of the AGN in our sample. The catalogue of \cite{Liu2016} includes L$_{bol}$ calculations. These have been derived by integrating the radiation directly produced by the accretion process, that is the thermal emission from the accretion disc and the hard X-ray radiation produced by inverse-Compton scattering of the soft disc photons by a hot corona (for more details see their Sect. 4.2). CIGALE also provides L$_{bol}$ measurements. \cite{Mountrichas2023b} compared the two L$_{bol}$ estimates and found that their distributions have a mean difference of 0.08\,dex with a standard deviation of 0.42\,dex. Following \cite{Mountrichas2023b}, we choose to use the L$_{bol}$ calculations of CIGALE. However, we note that using the L$_{bol}$ measurements from the \cite{Liu2016} catalogue does not affect our results and conclusions.

The n$_{Edd}$ is defined as the ratio of the bolometric luminosity, L$_{bol}$, and the Eddington luminosity, L$_{Edd}$. L$_{Edd}$ is the maximum luminosity that can emitted by the AGN and is determined by the balance between the radiation pressure and the gravitational force exerted by the black hole (L$_{Edd}=1.26\times 10^{38}\,M_{BH}/M_\odot$\,$\rm erg\,s^{-1}$). In our analysis, we use n$_{Edd}$ measurements derived using the L$_{bol}$ calculations from CIGALE, as opposed to those available in the \cite{Liu2016} catalogue. Nevertheless, this choice does not affect our results.

The $\lambda_{sBHAR}$ is the rate of the accretion onto the SMBH relative to the M$_*$ of the host galaxy. It is often used as a proxy of the Eddington ratio, in particular when black hole measurements are not available. For the calculation of $\lambda_{sBHAR}$ the following expression is used:
\begin{equation}
\lambda_{sBHAR}=\rm \frac{k_{bol}\,L_{X,2-10\,keV}}{1.26\times10^{38}\,erg\,s^{-1}\times0.002\frac{M_{*}}{M_\odot}},   
\label{eqn_lambda}
\end{equation}
where $\rm k_{bol}$ is a bolometric correction factor, that converts the $2-10$\,keV X-ray luminosity to AGN bolometric luminosity. For our sample, L$_{bol}$ measurements are already available, as described earlier in this section, and thus a bolometric correction is not required. Nevertheless, we choose to use equation \ref{eqn_lambda} for the calculation of $\lambda_{sBHAR}$, as it is the most common method to calculate $\lambda_{sBHAR}$ and it also will facilitate a direct comparison with the SFR$_{norm}-\lambda_{sBHAR}$ measurements of our previous studies \citep{Mountrichas2021c, Mountrichas2022a}. For the same reasons, instead of the M$_{BH}$ measurements that are available for our sources, we choose to use the redshift-independent scaling relation between M$_{BH}$ and bulge mass, M$_{bulge}$, of \cite{Marconi2003} with the assumption that the M$_{bulge}$ can be approximated by the M$_*$. Specifically, we use M$_{BH}=0.002$\,M$_{bulge}$.  Finally, for $\rm k_{bol}$, we adopt the value of $\rm k_{bol}=25$. This value is used in many studies \citep[e.g.,][]{Elvis1994, Georgakakis2017, Aird2018, Mountrichas2021c, Mountrichas2022a}. Lower values have also be used \citep[e.g., $\rm k_{bol}=22.4$ in][]{Yang2017}, as well as luminosity dependent bolometric corrections \citep[e.g.,][]{Hopkins2007a, Lusso2012}. In Sect. \ref{sec_lambda_vs_eddratio}, we examine how good these approximations are and what is their effect on the calculation of $\lambda_{sBHAR}$.

\section{Final samples}
\label{sec_final_samples}

In this section, we describe the criteria we apply to compile the final dataset of X-ray sources, drawn from the XMM-XXL catalogue (Sect. \ref{sec_data_xxl}) and the final control sample of non-AGN galaxies, drawn from the VIPERS survey (Sect. \ref{sec_data_vipers}).

\subsection{The final X-ray dataset}

We need to use only sources (X-ray and non-AGN galaxies) that have the most reliable M$_*$ and SFR measurements. For that purpose, for the X-ray sources, we use the final sample presented in \cite{Mountrichas2023b}. A detailed description of the photometric and reliability criteria that have been applied is provided in Sect. 2.4 of that study. In brief, we require our sources to have measurements in the following photometric bands: $u$, $g$, $r$, $i$, $z$, J, H, K, W1, W2 and W4, where W1, W2 and W4 are the WISE photometric bands at 3.4, 4.6 and 22\,$\mu$m. To exclude sources with bad SED fits and unreliable host galaxy measurements, a  reduced $\chi ^2$  threshold of $\chi ^2_r <5$ has been imposed \citep[e.g.][]{Masoura2018, Buat2021}. We also exclude systems for which CIGALE could not constrain the parameters of interest (SFR, M$_*$). Towards this end,  the two values that CIGALE provides for each estimated galaxy property are used. One value corresponds to the best model and the other value (bayes) is the likelihood-weighted mean value. A large difference between the two calculations suggests a complex likelihood distribution and important uncertainties. We therefore only include in our analysis sources with $\rm \frac{1}{5}\leq \frac{SFR_{best}}{SFR_{bayes}} \leq 5$ and $\rm \frac{1}{5}\leq \frac{M_{*, best}}{M_{*, bayes}} \leq 5$, where SFR$\rm _{best}$ and  M$\rm _{*, best}$ are the best-fit values of SFR and M$_*$, respectively and SFR$\rm _{bayes}$ and M$\rm _{*, bayes}$ are the Bayesian values estimated by CIGALE. 687 broad-line, X-ray AGN with spectroscopic redshifts meet the above requirements and also have available M$_{BH}$ measurements in the catalogue of \cite{Liu2016}. 

We then restrict the redshift range of the X-ray dataset to match that of the galaxy control sample (i.e., the VIPERS survey, $\rm 0.5\leq z\leq 1.2$). 240 AGN meet this requirement. In \cite{Mountrichas2021c, Mountrichas2022a, Mountrichas2022b}, we found that the SFR$_{norm}$-L$_X$ relation depends on the M$_*$ range probed by the sources. Specifically a flat SFR$_{norm}$-L$_X$ relation was found for the least and most massive systems ($\rm log\,[M_*(M_\odot)]<10.5$ and $\rm log\,[M_*(M_\odot)]>11.5$), with SFR$_{norm}\sim 1$. Albeit, for intermediate stellar masses ($\rm 10.5<log\,[M_*(M_\odot)]<11.5$) SFR$_{norm}$ was found to be $\leq 1$ at low-to-moderate L$_X$ ($\rm log\,[L_{X,2-10keV}(erg\,s^{-1})]<44$) whereas at higher L$_X$, SFR$_{norm}>1$  \citep[e.g., see Fig. 5 in][]{Mountrichas2022b}. Therefore, in this study, we restrict the analysis to those sources with $\rm 10.5<log\,[M_*(M_\odot)]<11.5$. Within this M$_*$ range both of our datasets are also mass complete \citep{Davidzon2013, Mountrichas2023}, as it required for the calculation of SFR$_{norm}$.


\begin{figure*}
\centering
  \includegraphics[width=0.85\columnwidth, height=6.2cm]{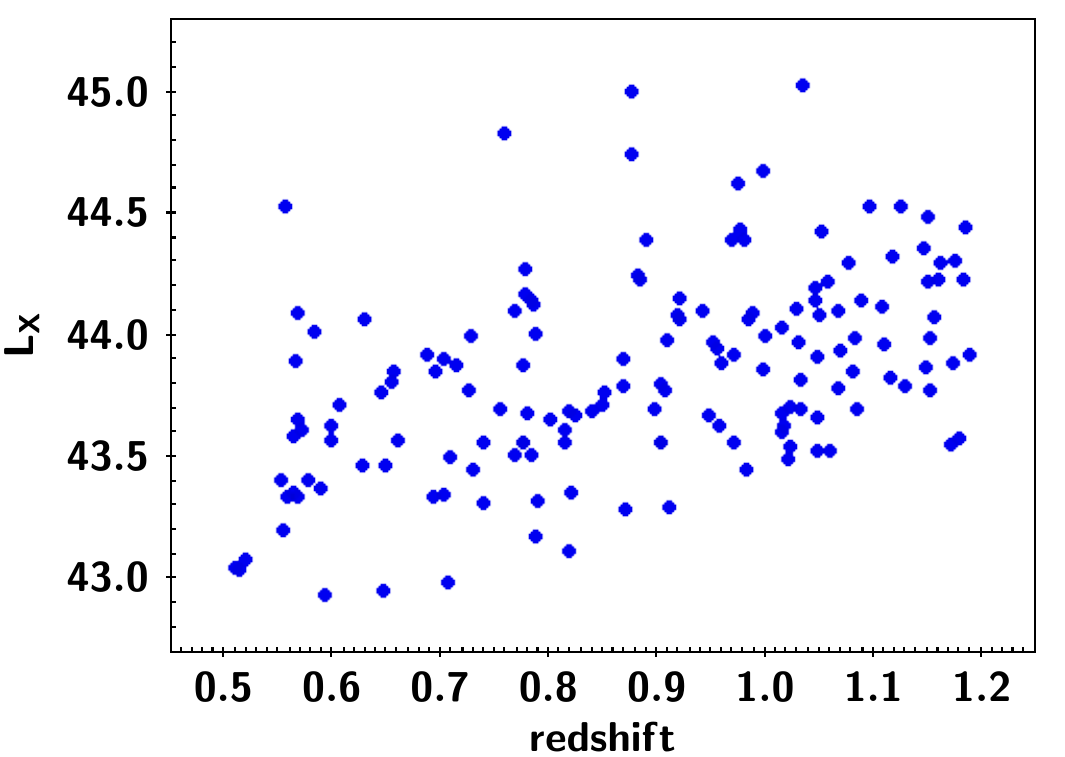}   
    \includegraphics[width=0.85\columnwidth, height=6.2cm]{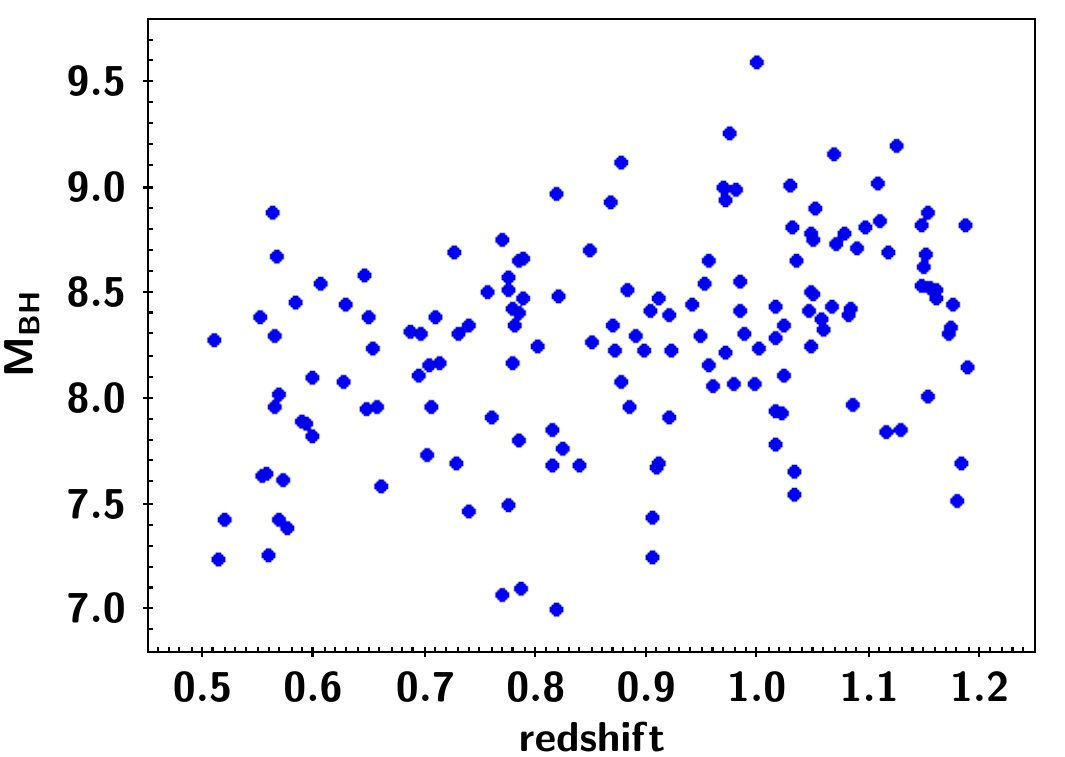} 
  \caption{L$_X$ (left panel) and M$_{BH}$ (right panel) as a function of redshift, for the 122 X-ray AGN used in our analysis.}
  \label{fig_lx_mbh_redz}
\end{figure*}

Following previous studies that examined the impact of the AGN feedback on their host galaxies, by calculating SFR$_{norm}$ using only star-forming systems \citep[e.g.,][]{Mullaney2015, Masoura2018, Mountrichas2021c}, we exclude from our sources quiescent (Q) systems. To identify Q galaxies, we use the distribution of the specific SFR ($\rm sSFR=\frac{SFR}{M_*}$) measurements of the galaxy control sample \citep[i.e., similarly to][]{Mountrichas2021c, Mountrichas2022a, Mountrichas2022b}. \cite{Mountrichas2023}, applied this methodology on sources in the XMM-XXL field to classify galaxies as Q. From their subset of Q sources, 19 are among our 178 AGN. Their exclusion results in 159 X-ray systems. We note that the inclusion of the 19 AGN hosted by Q systems in our analysis, does not affect our overall results and conclusions.

Since the galaxy control sample used in this study is smaller compared to those used in our previous works (see next section), we apply a final criterion to ensure that the SFR$_{norm}$ calculations of each AGN that is included in our analysis is robust. That is, we only use AGN that their SFR$_{norm}$ has been calculated by matching the X-ray sources with at least 300 sources in the galaxy control sample. Increasing this threshold reduces significantly the size of the X-ray dataset, while at lower values the scatter of our measurements is higher. 122 X-ray AGN fulfil all the aforementioned criteria. Their L$_X$ and M$_{BH}$ as a function of redshift are presented in Figure \ref{fig_lx_mbh_redz}.


\subsection{The final galaxy control sample}

For the galaxy control sample, we apply the same photometric selection criteria and reliability requirements that we applied for the X-ray AGN sample. In addition, we exclude sources that are included in the X-ray catalogue and we identify and reject non-X-ray AGN systems. Specifically, we use the CIGALE measurements and exclude sources with frac$\rm _{AGN}$ > 0.2, consistently with our previous studies \citep{Mountrichas2021c, Mountrichas2022a, Mountrichas2022b}. frac$\rm _{AGN}$ is the fraction of the total IR emission coming from the AGN. This excludes $\sim 60\%$ of the sources in the galaxy reference catalogue. This fraction is in line with our previous studies. A detailed analysis of the frac$\rm _{AGN}$ criterion is provided in Sect. 3.3 in \cite{Mountrichas2022b}. There are 3622 galaxies that fulfil all the aforementioned requirements. Finally, we exclude quiescent galaxies following the process described in the previous section. There are 3371 galaxies that remain and these are the sources in our control sample that we include in the analysis.


\begin{table}
\caption{p-values from the correlation analysis we apply for the four SMBH properties used in our analysis.}
\centering
\setlength{\tabcolsep}{2.mm}
\begin{tabular}{cccc}
relation& Pearson & Spearman & Kendall  \\
 \hline
 $M_{BH}$-L$_X$ &  $2.1\times 10^{-11}$ & $3.4\times 10^{-11}$ & $6.5\times 10^{-11}$ \\
 n$_{Edd}$-L$_X$ &  0.40 & 0.59 & 0.51   \\
 $\lambda _{sBHAR}$-L$_X$ & $2.9\times 10^{-15}$ & $2.0\times 10^{-14}$  &  $2.6\times 10^{-13}$ \\
 n$_{Edd}$-M$_{BH}$ &  $4.7\times 10^{-15}$  & $6.5\times 10^{-14}$   &  $1.1\times 10^{-12}$   \\
 $\lambda _{sBHAR}$-M$_{BH}$ &  $6.6\times 10^{-6}$  & $6.7\times 10^{-6}$   &  $5.2\times 10^{-6}$   \\
  $\lambda _{sBHAR}$-n$_{Edd}$ & $6.1\times 10^{-7}$ & $1.4\times 10^{-5}$  &  $1.3\times 10^{-5}$ \\
  \hline
\label{table_smbh_correlation}
\end{tabular}
\end{table}

\begin{figure*}
\centering
  \includegraphics[height=17cm]{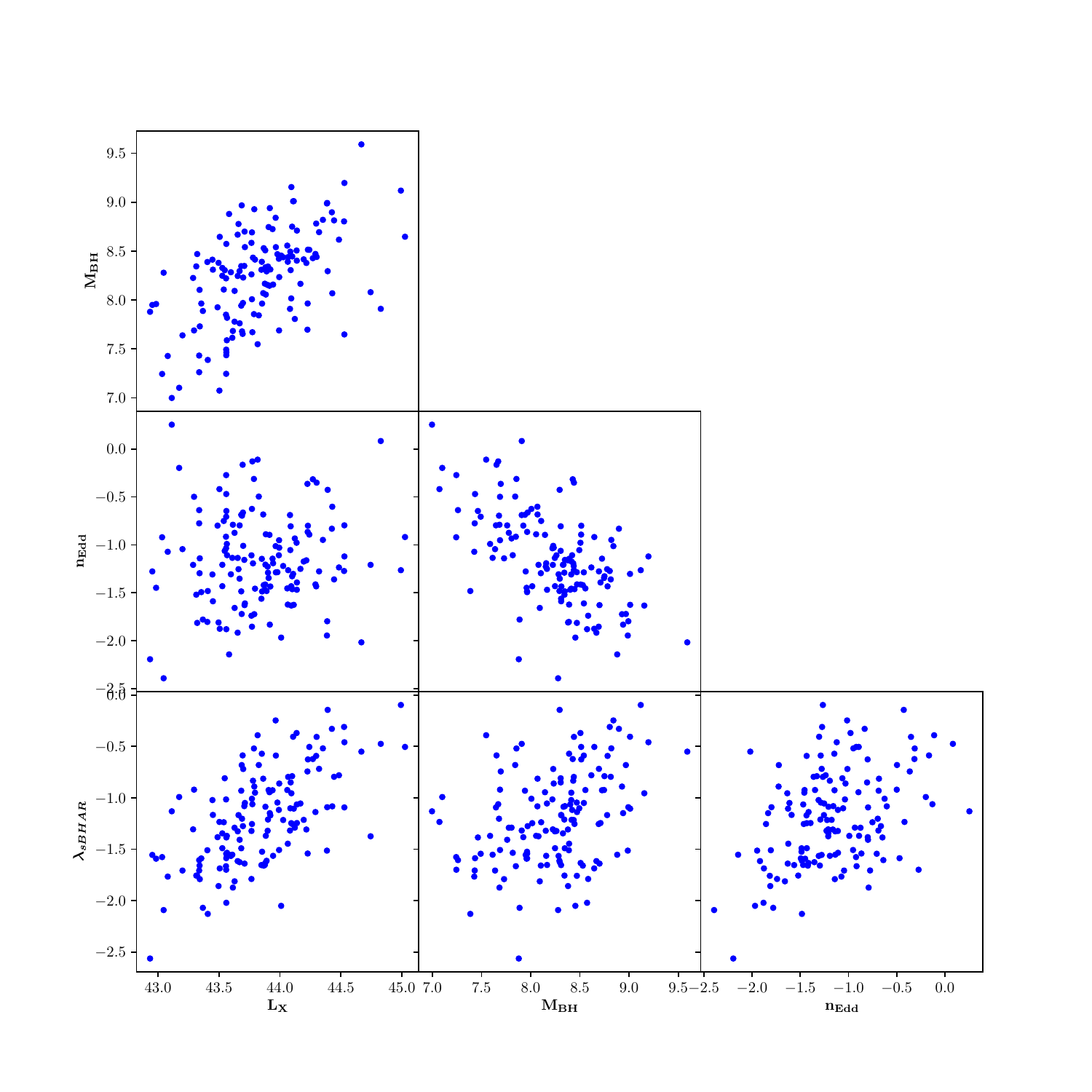}   
  \caption{Correlations among the four SMBH properties used in our study. Specifically, we present the correlations among the M$_{BH}$, the L$_X$, the specific black hole accretion rate ($\lambda _{sBHAR} \propto \frac{L_X}{M_*}$) and the Eddington ratio (n$_{Edd} \propto \frac{L_{bol}}{M_{BH}}$). The p-values from the correlation analysis are shown in Table \ref{table_smbh_correlation}.}
  \label{fig_smbh_properties}
\end{figure*}

\section{Results and Discussion}
\label{sec_results}

We compare the SFR of AGN and non-AGN galaxies as a function of various black hole properties. Specifically, we study SFR$_{norm}$ as a function of L$_X$, M$_{BH}$, n$_{Edd}$ and $\lambda _{sBHAR}$. Fig. \ref{fig_smbh_properties}, presents the four SMBH properties for the final X-ray dataset. We also apply three correlation statistics, one parametric (Pearson) and two non-parametric statistics (Spearman and Kendall), to quantify the correlations among them. The p-values are presented in Table \ref{table_smbh_correlation}. All parameters are strongly correlated with each other with the exception of the n$_{edd}-$L$_X$.

\begin{figure*}
\centering
  \includegraphics[width=0.85\columnwidth, height=6.2cm]{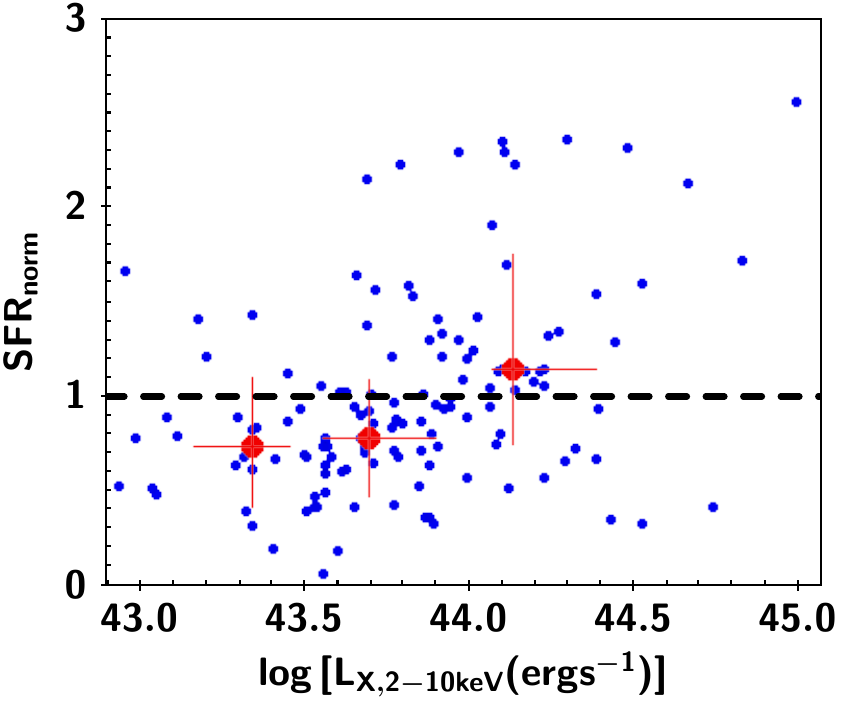}   
  \includegraphics[width=0.85\columnwidth, height=6.2cm]{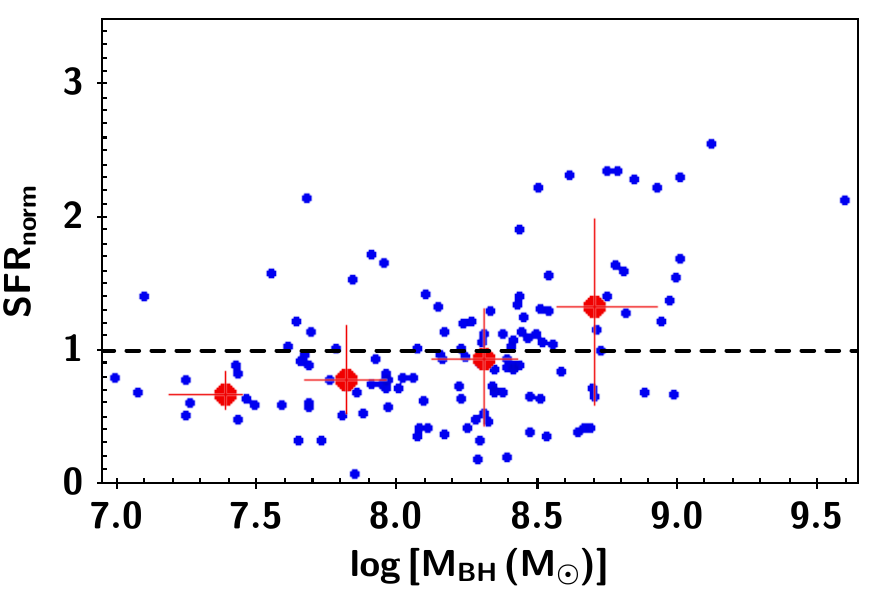} 
  \includegraphics[width=0.85\columnwidth, height=6.2cm]{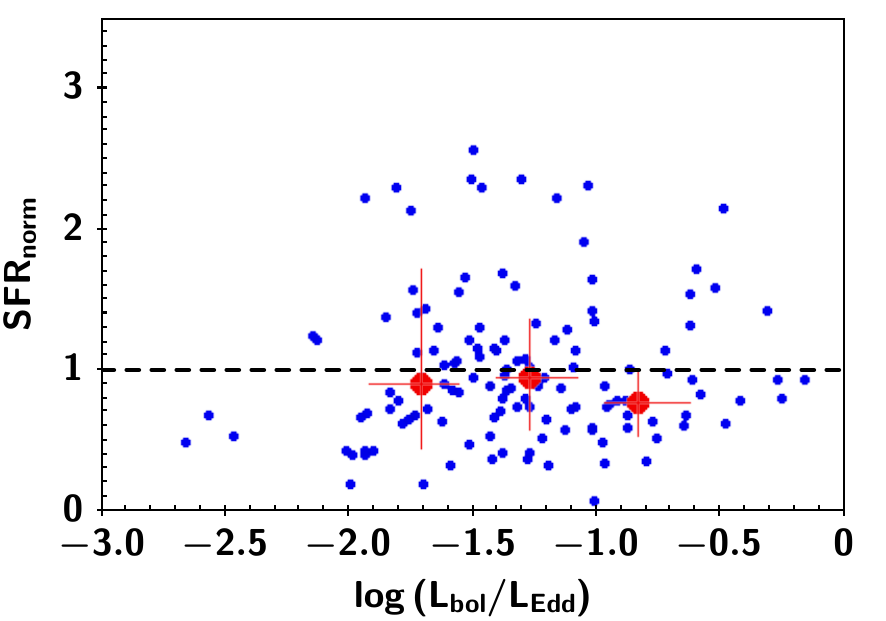}
    \includegraphics[width=0.85\columnwidth, height=6.2cm]{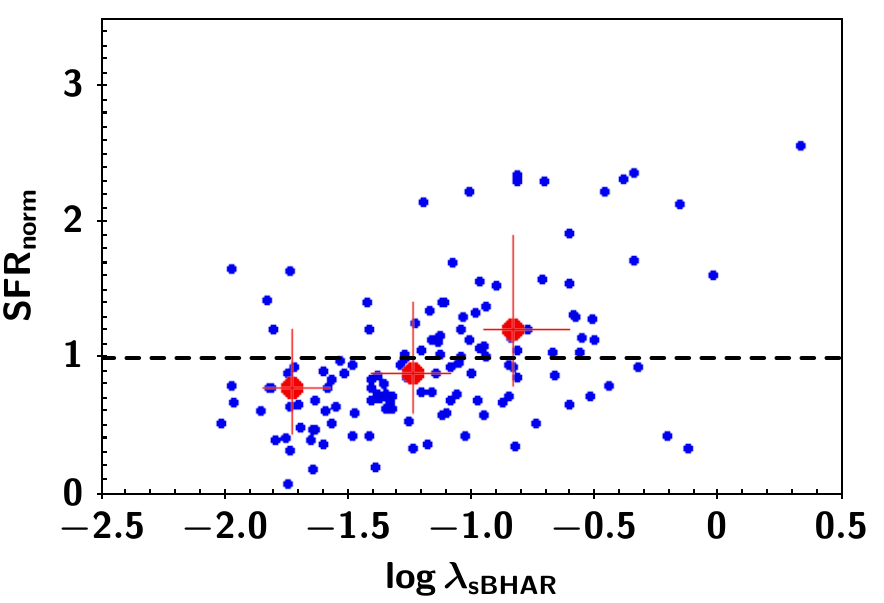}
  \caption{SFR$_{norm}$ as a function of SMBH properties. The SFR$_{norm}$ parameter as a function of L$_X$ (top, left panel), M$_{BH}$ (top, right panel), Eddington ratio (bottom, left panel) and $\lambda _{sBHAR}$ (bottom, right panel) are presented.}
  \label{fig_sfrnorm}
\end{figure*}

\subsection{SFR$_{norm}$ as a function of X-ray luminosity}

First, we examine SFR$_{norm}$ as a function of L$_X$. The results are shown in the left, top panel of Fig. \ref{fig_sfrnorm}. The small, blue circles present the measurements for individual AGN, while the large, red circles show the binned results. For the latter, the measurements are grouped in bins of L$_X$ of size 0.5\,dex. The errors presented are 1\,$\sigma$ errors, calculated via bootstrap resampling \citep[e.g.,][]{Loh2008}. We find that the SFR of AGN is lower or at most equal to that of non-AGN galaxies (SFR$_{norm}\leq 1$) at low and moderate L$_X$ ($\rm log\,[L_{X,2-10keV}(ergs^{-1})]\leq 44$) and increases at higher L$_X$, in agreement with previous studies \citep{Mountrichas2021c, Mountrichas2022a, Mountrichas2022b}.

The p-values from the three correlation statistics we use to calculate the correlation between SFR$_{norm}$ and L$_X$, are presented in Table \ref{table_correlation}. The results indicate a strong correlation between the two parameters, independent of the statistical method applied.


\subsection{SFR$_{norm}$ as a function of black hole mass}

In a recent study, \cite{Piotrowska2022}, analyzed three cosmological hydrodynamical simulations (Eagle, Illustris and IllustrisTNG), by utilizing  Random Forest classification. They searched for the most effective parameter to separate star-forming and quenched galaxies, in the local universe. They considered stellar mass, dark matter halo mass, black hole accretion rate and black hole mass in their investigation. Their analysis showed that black hole mass was the most predictive parameter of galaxy quenching. \cite{Bluck2023}, extended these results from the local universe to cosmic noon. These findings suggest that the cumulative impact of AGN feedback on a galaxy is encapsulated in the mass of the supermassive black hole and not in the X-ray luminosity, which is a proxy of the current accretion rate. 

Hence, here we examine the SFR$_{norm}$ as a function of black hole mass. Our goal is to examine if SFR$_{norm}$ and M$_{BH}$ are correlated and compare their correlation with that between SFR$_{norm}$ and L$_X$. The top, right panel of Fig. \ref{fig_sfrnorm} presents the SFR$_{norm}$ as a function of M$_{BH}$. The results show that SFR$_{norm}$ increases with M$_{BH}$ on the full range of black hole masses spanned by our dataset. Specifically, in galaxies that host AGN with low M$_{BH}$ ($\rm log\,[M_{BH}\,(M_\odot)]<8$) their SFR is lower or equal to the SFR of non-AGN systems. AGN with more massive black holes ($\rm log\,[M_{BH}\,(M_\odot)]> 8.5$) live in galaxies that their SFR is enhanced compared to non-AGN. The correlation analysis (Table \ref{table_correlation}) suggests a strong correlation between SFR$_{norm}$ and M$_{BH}$. 

We also split our datasets into two redshift bins, using a threshold at $\rm z=0.9$ and repeat the correlation analysis. The choice of the redshift cut is twofold. Primarily, it aligns with the median redshift of the AGN sample. Furthermore, this redshift value corresponds to the redshift at which different spectral lines have been used for the calculation of M$_{BH}$ (see Sect. \ref{sec_mbh}). The results are presented in Tables \ref{table_correlation_lowz} and \ref{table_correlation_highz}. The same trends are observed with those using sources in the full redshift interval, that is a strong correlation is found between SFR$_{norm}$ and M$_{BH}$ in both redshift ranges. However, this correlation appears less strong in the lowest redshift interval compared to that found in the highest redshift bin. This could imply that the correlation between the two properties is, mainly, driven by massive M$_{BH}$ (M$_{BH}>\sim 10^{8.5}\,M_\odot$) that are poorly detected at $\rm z<0.9$ in the dataset used in our analysis (Fig. \ref{fig_lx_mbh_redz}). This interpretation is also supported by the strong correlation between L$_X$ and M$_{BH}$ (Fig. \ref{fig_smbh_properties}) combined with the results from previous studies that have shown that the SFR$_{norm}-$L$_X$ relation is nearly flat at L$_X<10^{44}$\,erg/s and shows a positive correlation only at higher L$_X$ \citep{Mountrichas2021c, Mountrichas2022a, Mountrichas2022b}.

A comparison of the p-values with those in the previous section, shows that the correlation between SFR$_{norm}$ and M$_{BH}$ is similar to that between SFR$_{norm}$ and L$_X$. Subsequently, we explore whether this observation holds when considering the associated uncertainties of L$_X$ and M$_{BH}$. For that purpose, we utilize the linmix module \citep{Kelly2007} that performs linear regression between two parameters, by repeatedly perturbing the datapoints within their uncertainties. The p-values obtained are $3.2\times 10^{-5}$ and $7.6\times 10^{-4}$ for the SFR$_{norm}-$L$_X$ and SFR$_{norm}-$M$_{BH}$, respectively. These findings suggest, that despite accounting for uncertainties in L$_X$ and M$_{BH}$ measurements, there exists a robust correlation between these two properties and SFR$_{norm}$ and that the two correlations are indeed similar.

As shown in Fig. \ref{fig_smbh_properties} and Table \ref{table_smbh_correlation}, L$_X$ and M$_{BH}$ are strongly correlated. To investigate further the correlation among SFR$_{norm}$, L$_X$ and M$_{BH}$, we perform a partial-correlation analysis (PCOR). PCOR measures the correlation between two variables while controlling for the effects of a third \citep[e.g.][]{Lanzuisi2017, Yang2017, Mountrichas2022a}. We use one parametric statistic (Pearson) and one non-parametric statistic (Spearman). Table \ref{table_pcor} lists the results of the p-values. Regardless of the parametric statistic of choice, p-values for the SFR$_{norm}-$M$_{BH}$ relation are smaller compared to the corresponding p-values for the SFR$_{norm}-$L$_X$ relation. This implies that the correlation between SFR$_{norm}$ and M$_{BH}$ is more robust compared to that with L$_X$, even when factoring in the existing correlation between M$_{BH}$ and L$_X$. This deduction remains valid even when we partition the dataset into two redshift bins, specifically at $\rm z=0.9$.

\cite{Mountrichas2022a} applied PCOR analysis on sources in the COSMOS field and found that SFR$_{norm}$ is correlated stronger with M$_*$ than with L$_X$. \cite{Yang2017} used galaxies in the CANDELS/GOODS-South field and examined the correlation between the black hole accretion rate (BHAR; which is measured directly from the L$_X$), SFR and M$_*$. They found that the BHAR is linked mainly to M$_*$ rather than SFR. There is also a well known correlation between the M$_*$ and the M$_{BH}$ \citep[e.g.,][]{Merloni2010, Sun2015, Suh2020, Setoguchi2021, Poitevineau2023}. Recently, \cite{Mountrichas2023b} reported such a correlation between M$_{BH}$ and M$_*$ using AGN in the XMM-XXL field, that is the same X-ray dataset used in this work. We apply a PCOR analysis, this time among SFR$_{norm}$, M$_{BH}$ and M$_*$. The results presented in Table \ref{table_pcor2} (top two lines) suggest that SFR$_{norm}$ is linked more to M$_{BH}$ than M$_*$. However, we note that, for the reasons mentioned in Sect. \ref{sec_final_samples}, our datasets have been restricted to a relatively narrow M$_*$ range ($\rm 10.5<log\,[M_*(M_\odot)]<11.5$). Therefore, although the M$_{BH}$ parameter spans $\sim 2.5$ orders of magnitude, M$_*$ spans only an order of magnitude in our samples. 

To increase the M$_*$ range that our sources probe, we lift the M$_*$ requirement. There are 209 AGN and 4454 galaxies within $\rm 10<log\,[M_*(M_\odot)]<12$. Using these two subsets we calculate the SFR$_{norm}$ for the 240 AGN and, then, we apply a PCOR analysis among SFR$_{norm}$, M$_{BH}$ and M$_*$. The results are presented in the two bottom lines of Table \ref{table_pcor2}. The p-values of the non-parametric statistic (Spearman) are similar, however, the p-value using the parametric statistic (Pearson) are lower for the SFR$_{norm}-$M$_{BH}$, suggesting that the correlation between SFR$_{norm}-$M$_{BH}$ is stronger than the correlation between SFR$_{norm}-$M$_*$. We note that these results should be taken with caution since our samples are not mass complete in the full M$_*$ range that is considered in this exercise and specifically within $\rm 10.0<log\,[M_*(M_\odot)]<10.5$.

Overall, we conclude that SFR$_{norm}$ is mostly linked to M$_{BH}$ rather than L$_X$. Our results also suggest that the SFR$_{norm}-$M$_*$ correlation is due to the underlying M$_*-$M$_{BH}$. The picture that emerges corroborates the idea that the M$_{BH}$ is a more robust tracer of AGN feedback compared to the instantaneous activity of the SMBH - represented by L$_X$ - and as such M$_{BH}$ is a better predictive parameter of the changes of the SFR of the host galaxy, as theoretical studies have also suggested \citep{Piotrowska2022, Bluck2023}. Our results are also in line with the aforementioned studies regarding the negative AGN feedback they report, at least up to M$_{BH}\sim 10^{8.5}\,M_\odot$ (i.e., SFR$_{norm}<1$). The increase of SFR$_{norm}$ we detect in our results, suggest that this negative feedback may become less impactful on the SFR of the host galaxy, as we transition to systems with more massive SMBHs. These studies have additionally shown that the fraction of quenched galaxies increases with M$_{BH}$. To investigate this claim, we would need to examine the fraction of quiescent systems as a function of M$_{BH}$, in our dataset. However, the small sample size used in our analysis and the low number of quiescent systems included, do not allow for such an investigation.

\begin{table}
\caption{p-values of correlation analysis, using sources with $\rm 0.5 \leq z \leq 1.2$.}
\centering
\setlength{\tabcolsep}{2.mm}
\begin{tabular}{cccc}
relation& Pearson & Spearman & Kendall  \\
  \hline
SFR$_{norm}$-L$_X$ & $3.1\times 10^{-6}$ & $2.9\times 10^{-7}$ & $1.4\times 10^{-7}$\\  
SFR$_{norm}$-M$_{BH}$ & $4.0\times 10^{-7}$ & $3.3\times 10^{-7}$ & $2.9\times 10^{-7}$\\
SFR$_{norm}$-n$_{Edd}$ & 0.87 & 0.56 & 0.58 \\
SFR$_{norm}$-$\lambda_{sBHAR}$ &$6.3\times 10^{-5}$ &$5.1\times 10^{-6}$ & $3.0\times 10^{-6}$\\
  \hline
\label{table_correlation}
\end{tabular}
\end{table}

\begin{table}
\caption{p-values of correlation analysis, using sources with $\rm 0.5 \leq z \leq 0.9$.}
\centering
\setlength{\tabcolsep}{2.mm}
\begin{tabular}{cccc}
relation& Pearson & Spearman & Kendall  \\
  \hline
SFR$_{norm}$-L$_X$ & $8.0\times 10^{-3}$ & $5.0\times 10^{-3}$ & $4.1\times 10^{-3}$\\  
SFR$_{norm}$-M$_{BH}$ & $2.1\times 10^{-3}$ & $4.2\times 10^{-3}$& $4.4\times 10^{-3}$\\
SFR$_{norm}$-n$_{Edd}$ & 0.75 & 0.48 & 0.48 \\
SFR$_{norm}$-$\lambda_{sBHAR}$ &$7.0\times 10^{-2}$ &$9.1\times 10^{-3}$ &$1.2\times 10^{-2}$\\
  \hline
\label{table_correlation_lowz}
\end{tabular}
\end{table}

\begin{table}
\caption{p-values of correlation analysis, using sources with $\rm 0.9 < z \leq 1.2$.}
\centering
\setlength{\tabcolsep}{2.mm}
\begin{tabular}{cccc}
relation& Pearson & Spearman & Kendall  \\
  \hline
SFR$_{norm}$-L$_X$ & $6.9\times 10^{-7}$ & $4.6\times 10^{-7}$ & $1.1\times 10^{-6}$\\  
SFR$_{norm}$-M$_{BH}$ & $1.7\times 10^{-7}$ & $2.6\times 10^{-7}$ & $1.7\times 10^{-6}$\\
SFR$_{norm}$-n$_{Edd}$ & 0.82 & 0.32 & 0.31 \\
SFR$_{norm}$-$\lambda_{sBHAR}$ &$1.4\times 10^{-6}$ &$8.4\times 10^{-7}$ & $2.8\times 10^{-6}$\\
  \hline
\label{table_correlation_highz}
\end{tabular}
\end{table}

\subsection{SFR$_{norm}$ as a function of Eddington ratio and specific black hole accretion rate}

In this section, we investigate the correlation between SFR$_{norm}$ and two other SMBH properties, that represent the instantaneous AGN activity. Specifically, we study the relation between SFR$_{norm}-$n$_{Edd}$ and SFR$_{norm}-\lambda _{sBHAR}$. We also examine whether $\lambda _{sBHAR}$ is a good proxy of the n$_{Edd}$.

\subsubsection{SFR$_{norm}$ as a function of Eddington ratio}

The Eddington ratio provides another important property of the SMBH. \cite{Setoguchi2021} used 85 moderately luminous ($\rm log\,L_{bol}\sim 44.5-46.5\,erg\,s^{-1}$) in the Subaru/XMM-Newton Deep Field (SXDF) and found a strong correlation between the SFR of AGN and n$_{Edd}$ (correlation coefficient: $r=0.62$). Recently, \cite{Georgantopoulos2023} studied the stellar populations of obscured and unobscured AGN at $\rm 0.6<z<1.0$. Based on their analysis, the stellar age of both AGN types increases at lower Eddington ratio values (see the bottom left panel of their Fig. 4 and the top, right panel of their Fig. 11).

The bottom, left panel of Fig. \ref{fig_sfrnorm}, presents our calculations for SFR$_{norm}$ as a function of the Eddington ratio. SFR$_{norm}$ remains roughly constant regardless of the value of n$_{Edd}$. This is confirmed by the results of the correlation analysis, shown in Table  \ref{table_correlation} (see also Tables \ref{table_correlation_lowz} and \ref{table_correlation_highz} for different redshift intervals). This nearly flat SFR$_{norm}-$n$_{Edd}$ relation can be explained by the correlations among the M$_{BH}$, L$_X$ and n$_{Edd}$, presented in Fig. \ref{fig_smbh_properties}. There is a strong anti-correlation between n$_{Edd}$ and M$_{BH}$, but a positive correlation between n$_{Edd}$ and L$_X$, while a strong positive correlation is detected between M$_{BH}$ and L$_X$. We note, that, when we examine the relation between the SFR of AGN and n$_{Edd}$, we find a (strong) correlation ($r=0.54$), similar to that found by \cite{Setoguchi2021}. 


\begin{table}
\caption{p-values of partial correlation analysis, among SFR$_{norm}$, L$_X$ and M$_{BH}$.}
\centering
\setlength{\tabcolsep}{2.mm}
\begin{tabular}{ccc}
& Pearson & Spearman  \\
  \hline
SFR$_{norm}$-L$_X$ & 0.056 & 0.016 \\  
SFR$_{norm}$-M$_{BH}$ & 7$\times 10^{-5}$ & 9$\times 10^{-5}$\\
  \hline
\label{table_pcor}
\end{tabular}
\end{table}

\begin{table}
\caption{p-values of partial correlation analysis, among SFR$_{norm}$, M$_*$ and M$_{BH}$.}
\centering
\setlength{\tabcolsep}{2.mm}
\begin{tabular}{ccc}
& Pearson & Spearman  \\
  \hline
SFR$_{norm}$-M$_*$ & 0.515 & 0.0068 \\  
SFR$_{norm}$-M$_{BH}$ & 1.6$\times 10^{-8}$ & 2.5$\times 10^{-9}$\\
SFR$_{norm}$-M$_*$ (ext) & 0.027 & 2$\times 10^{-6}$ \\  
SFR$_{norm}$-M$_{BH}$ (ext) & 1.1$\times 10^{-5}$ & 5$\times 10^{-6}$\\
  \hline
\label{table_pcor2}
\end{tabular}
\tablefoot{The top two lines present the results using sources within $\rm 10.5<log\,[M_*(M_\odot)]<11.5$. The bottom two lines present the results within $\rm 10<log\,[M_*(M_\odot)]<12$}
\end{table}

\subsubsection{SFR$_{norm}$ as a function of specific black hole accretion rate}
\label{sec_lambda}

The specific black hole accretion rate is often used as a proxy of the Eddington ratio. Previous studies found an increase of the SFR$_{norm}$ with $\lambda_{sBHAR}$ \cite[Figures 10 and 11 in][respectively]{Mountrichas2021c, Mountrichas2022a}. \cite{Pouliasis2022}, used X-ray AGN in the COSMOS, XMM-XXL and eFEDS, at $\rm z>3.5$ and found that AGN that lie inside or above the main-sequence (i.e., SFR$_{norm}\geq 1$) have higher  $\lambda_{sBHAR}$ compared to X-ray sources that lie below the MS.

Our results, presented in the bottom, right panel of Fig. \ref{fig_sfrnorm} agree with these previous findings. Specifically, we observe an increase of SFR$_{norm}$ with $\lambda_{sBHAR}$. Application of correlation analysis shows that there is a strong correlation between the two parameters, albeit not as strong as the correlation found between SFR$_{norm}$-L$_X$ and SFR$_{norm}$-M$_{BH}$ (Tables \ref{table_correlation}, \ref{table_correlation_lowz} and \ref{table_correlation_highz}).

 \cite{Mountrichas2022a} examined the correlation between SFR$_{norm}$ and $\lambda_{sBHAR}$ using X-ray sources in the COSMOS field and compared their results with those using AGN in the Bo$\ddot{o}$tes, presented in \cite{Mountrichas2021c} \citep[see Fig. 11 and Table 5 in][]{Mountrichas2022a}. Although both datasets present a nearly, linear increase of the SFR$_{norm}$ with L$_X$, the amplitude of SFR$_{norm}$ differs for the same $\lambda_{sBHAR}$  values, for the two datasets. They attributed this difference to the different properties of the AGN from the two samples included in $\lambda_{sBHAR}$  bins of the same value. Specifically, COSMOS sources are less luminous and less massive than their Bo$\ddot{o}$tes counterparts in $\lambda_{sBHAR}$ bins of similar values. Therefore, if a dataset probes AGN within a large range of L$_X$ and M$_*$, this could increase the scatter of SFR$_{norm}$ for the same $\lambda_{sBHAR}$  values and thus weaken the correlation between SFR$_{norm}$ and $\lambda_{sBHAR}$, rendering $\lambda_{sBHAR}$  not a good parameter to study the impact of AGN feedback on the SFR of the host galaxy.


\begin{figure}
\centering
  \includegraphics[width=0.85\columnwidth, height=6.2cm]{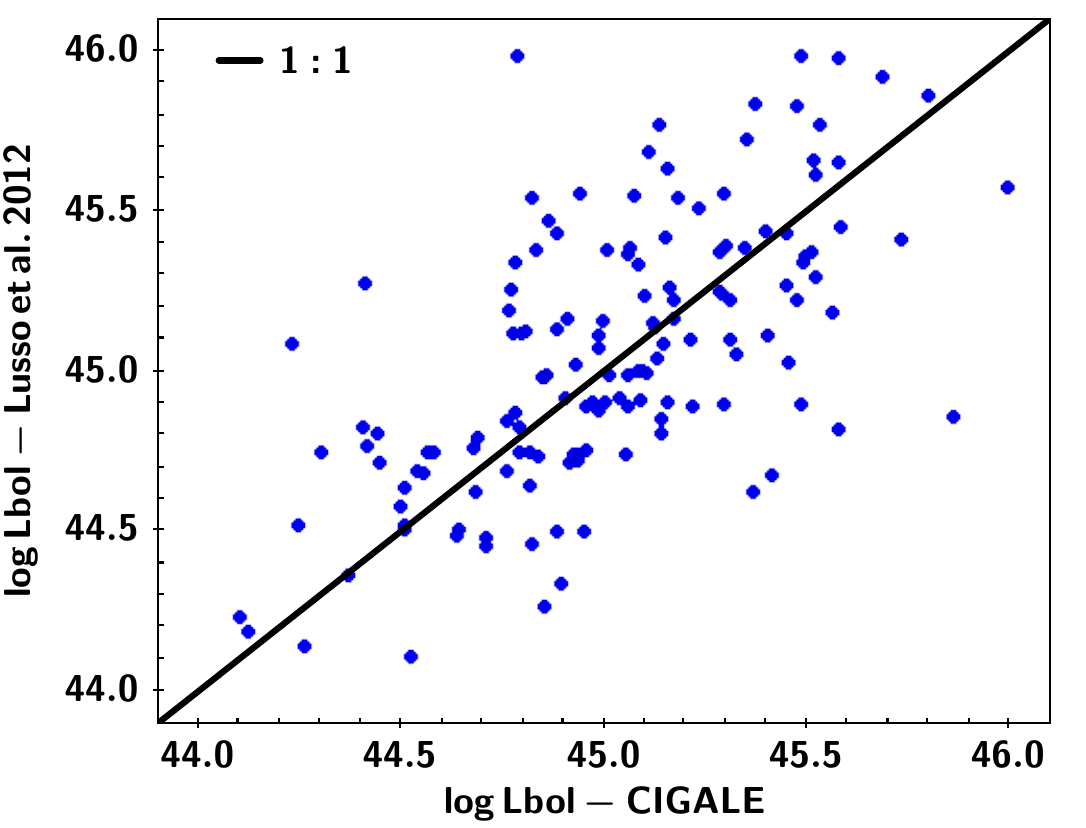}   
  \caption{Comparison of the L$_{bol}$ calculations of CIGALE with the L$_{bol}$ measurements using the formula derived in \cite{Lusso2012}. The two measurements are in very good agreement with a mean difference of 0.04\,dex. and a dispersion of 0.34.}
  \label{fig_kbol}
\end{figure}  

\subsubsection{Is $\lambda_{sBHAR}$ a good proxy of the Eddington ratio?}
\label{sec_lambda_vs_eddratio}

As mentioned in the previous section, $\lambda_{sBHAR}$ is often used as a proxy of n$_{Edd}$ on the basis that there is a linear relation between the M$_*$ and M$_{BH}$ and that L$_{bol}$ can be inferred by L$_X$. Prompted by the different relations found between SFR$_{norm}-$n$_{Edd}$ and SFR$_{norm}-\lambda _{sBHAR}$, we investigate this further.

\cite{Lopez2023} used X-ray selected AGN in the miniJPAS footprint and found, among others, that the Eddington ratio and  $\lambda_{sBHAR}$ have a difference of 0.6\,dex. They attributed this difference to the scatter on the M$_{BH}-$M$_*$ relation of their sources. The median value of n$_{Edd}$ of our sample, calculated using the L$_{bol}$ measurements of CIGALE, is n$_{Edd}=-1.26$, \citep[n$_{Edd}=-1.33$, using the values available in the][catalogue]{Liu2016}. The median value of $\lambda_{sBHAR}$, estimated using eqn \ref{eqn_lambda}, is $\lambda_{sBHAR}=-1.08$. Thus, we find a median difference of $\sim 0.25$ between n$_{Edd}$ and $\lambda_{sBHAR}$. Although this difference is lower than that reported by \cite{Lopez2023}, below we examine the cause of it. 


We re-calculate $\lambda_{sBHAR}$, using the L$_{bol}$ measurements from CIGALE (see Sect. \ref{sec_lbol_nedd_lambda}) instead of the product of $\rm k_{bol}\,L_X$. In this case, the median value of $\lambda_{sBHAR}$ is $-1.25$. This value is in excellent agreement with that of n$_{Edd}$ ($-1.26$), using for the calculation of the latter the L$_{bol}$ measurements from CIGALE. We also calculate $\lambda_{sBHAR}$ keeping the same numerator as in eqn \ref{eqn_lambda}, but using the M$_{BH}$ measurements available in our dataset instead of the M$_{BH}-$M$_*$ scaling relation. In this case, the median difference between the distributions of $\lambda_{sBHAR}$ and n$_{Edd}$ is $\sim 0.08$. We note that for the sources used in our analysis, the scaling relation between M$_{BH}$ and M$_*$ is, M$_{BH}\approx 0.003$\,M$_{*}$ \cite[see also Sect. 3.3 in][]{Mountrichas2023b}, which is in good agreement with the M$_{BH}=0.002$\,M$_{bulge}$ used in eqn \ref{eqn_lambda}. 

Therefore, the way the L$_{bol}$ is calculated seems to play an equally important role with the M$_{BH}-$M$_*$ scaling relation on the comparison between n$_{Edd}$ and $\lambda _{sBHAR}$, in our sample. The mean difference between the L$_{bol}$ calculated by CIGALE and the product of $\rm k_{bol}$\,L$_X$ is 0.24\,dex with a dispersion of 0.35. CIGALE measurements suggest a mean $\rm k_{bol}=14.8$ (i.e., for the two L$_{bol}$ measurements to have a mean difference of zero). Finally, we compare the L$_{bol}$ measurements of CIGALE with those using a luminosity dependent $k_{bol}$. Specifically, we use the prescription of \cite{Lusso2012}, using the values presented in their Table 2 for their spectroscopic, type-1 AGN. In this case, the two calculations are in very good agreement with a mean difference of 0.04\,dex and a dispersion of 0.34. Fig. \ref{fig_kbol} presents the comparison between the L$_{bol}$ measurements using the formula presented in \cite{Lusso2012} and CIGALE.


We conclude that caution has to be taken when $\lambda_{sBHAR}$ is used as a proxy of n$_{Edd}$, since the calculation of L$_{bol}$ and the scatter in the M$_{BH}-$M$_*$ scaling relation can cause (large) discrepancies between the estimated values of the two parameters.

\section{Conclusions}
\label{sec_conclusions}

We used 122 X-ray AGN in the XMM-XXL-N field and 3371 VIPERS galaxies, within redshift and stellar mass ranges of $\rm 0.5\leq z\leq 1.2$ and $\rm 10.5<log\,[M_*(M_\odot)]<11.5$, respectively. The X-ray sources probe luminosities within $\rm 43<log\,[L_{X,2-10keV}(ergs^{-1})]<45$. Both populations meet strict photometric selection criteria and various selection requirements to ensure that only sources with robust (host) galaxy measurements are included in the analysis. The latter have been calculated via SED fitting, using the CIGALE code. Using these datasets, we calculated the SFR$_{norm}$ parameter, to compare the SFR of AGN with the SFR of non-AGN galaxies, as a function of various black hole properties. Specifically, we examined the correlations of SFR$_{norm}$ with the L$_X$, M$_{BH}$, n$_{Edd}$ and $\lambda _{sBHAR}$. Our main results can be summarized as follows:

\begin{itemize}


\item[$\bullet$] AGN with low black hole masses ($\rm log\,(M_{BH}/M_*)<8$) have lower or at most equal SFR compared to that of non-AGN galaxies, while AGN with more massive black holes ($\rm log\,(M_{BH}/M_*)>8.5$) tend to live in galaxies with (mildly) enhanced SFR compared to non-AGN systems.

\item[$\bullet$] SFR$_{norm}$ strongly correlates with both L$_X$ and M$_{BH}$. However, the correlation between SFR$_{norm}-$M$_{BH}$ is stronger compared to the correlation between SFR$_{norm}-$L$_X$. Our results also suggest that M$_{BH}$ drives the correlation between SFR$_{norm}-$M$_*$ found in previous studies. 

\item[$\bullet$] We do not detect a significant correlation between SFR$_{norm}$ and Eddington ratio.

\item[$\bullet$] A correlation is found between SFR$_{norm}$ and specific black hole accretion rate. However, this correlation is weaker compared to that between SFR$_{norm}-$L$_X$ and SFR$_{norm}-$M$_{BH}$ and its scatter may increase for samples that span a wide range of L$_X$ and M$_*$.

\item[$\bullet$] The estimation of the AGN bolometric luminosity and the scatter of the M$_{BH}-$M$_*$ scaling relation, may cause discrepancies between the specific black hole accretion rate and the Eddington ratio measurements. Therefore, caution has to be taken when the former is used as a proxy for the latter.

\end{itemize}

The results suggest that there is a strong correlation between SFR$_{norm}$ and AGN activity, when the latter is represented by L$_X$, $\lambda _{sBHAR}$ and M$_{BH}$. A flat relation was only found between SFR$_{norm}$ and n$_{Edd}$, that can be interpreted as the net result of the different correlations (i.e., positive and negative) among n$_{edd}$, M$_{BH}$ and L$_X$ (Fig. \ref{fig_smbh_properties}). Based on our analysis, M$_{BH}$ is the most robust tracer of AGN feedback and the best predictive parameter of the changes of the SFR of the host galaxy.

\begin{acknowledgements}
This project has received funding from the European Union's Horizon 2020 research and innovation program under grant agreement no. 101004168, the XMM2ATHENA project.
The project has received funding from Excellence Initiative of Aix-Marseille University - AMIDEX, a French 'Investissements d'Avenir' programme.
This work was partially funded by the ANID BASAL project FB210003. MB acknowledges support from FONDECYT regular grant 1211000.
This research has made use of TOPCAT version 4.8 \citep{Taylor2005}.

\end{acknowledgements}

\bibliography{mybib}
\bibliographystyle{aa}

\end{document}